\documentclass[aps,superscriptaddress,hyperref]{revtex4}
\usepackage{graphicx,psfrag,amsmath,times,hhline}
\usepackage{epsf}
\usepackage{epsfig}
\newcommand{\be}{\begin{equation}}
\newcommand{\ee}{\end{equation}}
\newcommand{\br}{\begin{eqnarray}}
\newcommand{\er}{\end{eqnarray}}

\begin{document}

\title {
The Superposition Principle in Quantum Mechanics - did the rock enter the foundation surreptitiously?
}
\author{N.D.~Hari Dass}
\affiliation{Chennai Mathematical Institute,Chennai, India
.}
\affiliation{CQIQC, IISc, Bangalore, India
.}

\begin{abstract}
 The superposition principle forms the very backbone of quantum theory.
The resulting linear structure of quantum theory is structurally
so rigid that tampering with it may have serious, seemingly unphysical, consequences. 
This principle
has been succesful at even the highest available accelerator energies.
Is this aspect of quantum theory forever then?
The present work is an attempt to understand the attitude of the founding fathers, particularly of Bohr and Dirac,
towards this principle. The Heisenberg matrix mechanics on the one hand, and the Schr\"odinger wave mechanics
on the other, are critically examined to shed light as to how this principle entered the very foundations of quantum
theory.
\end{abstract}
\maketitle

\section{Introduction}
The superposition principle for \emph{quantum states}
 can be said to be the very bedrock of quantum theory.
To paraphrase Dirac,
 \emph{This principle forms the fundamental new idea of quantum mechanics and the basis for the departure from
classical theory}(Dirac in the first edition of his book \cite{diracbook1}, $\S 2$, p.2).
While commenting on the new laws required for the description of atomic phenomena, he emphasized that
\emph{One of the most fundamental and most drastic of these is {\it the Principle of superposition of states}}(see $\S 2$ p.4 
of \cite{diracbook2}). Dirac went as far as to assert: '\emph{One could proceed to build up the theory of quantum
mechanics on the basis of these ideas of superposition with the introduction of the minimum numbe of new assumptions necessary.}'(p.16 of 
$\S 6$ of \cite{diracbook1}).\\

Immediately after the discovery of matrix mechanics and wave-mechanics \cite{qorig}, while Heisenberg's uncertainty principle \cite{heisenunc}, Bohr's complimentarity principle \cite{bohrcomp}, and Born's 
probability rule \cite{bornprob}
paved the way for a lasting physical interpretation of quantum theory, an equally important parallel development was the work of Dirac \cite{diracbook1,diracbook2,diracpaper1,diracpaper2,diracpaper3} 
who focussed on the
nature of the \emph{quantum state}. This eventually culminated in the formulation of the \emph{principle of superposition of states}.
\subsection{The Superposition Principle for Quantum States}
Though it is customary to view the superposition principle within the mathematical framework of \emph{Hilbert Spaces}, it is 
instructive to recall its purely \emph{operational} meaning as elaborated by Dirac. He gives a very broad characterization of 
\emph{states} as the embodiment of the collection of all possible measurement outcomes. Then superposition of states according to him
is as follows: if A is a superposition of two or more states, say, B,C,.., every outcome of a measurement on A must also be a
possible outcome of the same measurement on any of B, C, ..(p.15 $\S 6$ of \cite{diracbook1}). Though this characterization of superposition may seem \emph{ad hoc},
the customary, Hilbert Space based view is completely equivalent to it. but the Dirac characterization has the advantage of being
purely operational and applicable even if there is no underlying Hilbert space structure.

 Being about superposition of \emph{states}, it is like \emph{no other} superposition principle in either
physics or mathematics.
 Examples of the latter are superposition of sound waves, of electromagnetic waves, of vectors etc..
 This was most emphatically stated by Dirac himself: '\emph{It is important to remember, however,
that {\it the superposition that occurs in quantum mechanics is of an essentially different nature from any occurring in the classical
theory}}' (The italics are Dirac's). 
 He further stated that \emph{The analogies are thus liable to be misleading}(p.11 $\S 3$ of \cite{diracbook1}, and p.14 $\S 4$ of 
\cite{diracbook2}).

In the current formulation of quantum theory, this principle is given a precise mathematical meaning 
through the \emph{Hilbert Space} formalism(actually one needs the \emph{density matrix} formalism for a more satisfactory description, but
that discussion is somewhat beyond the scope of this presentation). According to this, every physical state is representable by a \emph{family}
of vectors in a Hilbert space. A typical such vector is symbolically denoted by $|\rangle$. 
Vectors belonging to a given family differ only in phase.
This is the so called \emph{ray representation} of states.
If $P_1, P_2$ are two \emph{distinct} physical states meaning their rays are distinct, and if $|1\rangle$ belongs to the ray of $P_1$ and
$|2\rangle$ belongs to the ray of $P_2$,  the principle of superposition of states states that 
the \emph{complex linear superpositions}
\be
|\psi\rangle = \alpha|1\rangle + \beta |2\rangle
\ee
also represent \emph{quantum states} of the system.
\subsection{Illustration through an artistic example..}
 Imagine a \emph{rose atom} which exists in only two colour states, say, \emph{$|red\rangle$} and \emph{$|yellow\rangle$}.
 According to present day quantum mechanics, this means that if you observe \emph{colour}, the
outcome will be \emph{red} if the system was in state $|red\rangle$, and \emph{yellow} if the system was in $|yellow\rangle$. 
It should be emphasized that these rules of \emph{quantum measurements} are themselves intimately tied to the superposition principle.
Now, according to  the principle of superposition of states, linear combinations of the type (and in fact infinitely more)
\be
|good\rangle = \frac{1}{\sqrt{2}}\,\left\{|red\rangle+|yellow\rangle\right\}
\ee
and
\be
 |bad\rangle = \frac{1}{\sqrt{2}}\,\left\{|red\rangle-|yellow\rangle\right\}
\ee
also represent legitimate quantum states!

This immediately leads to the problematic question \emph{"What will be the outcome of a colour measurement on these new states?"}
 \emph{If} the measurement scheme follows the standard lore, the outcome can only be red or yellow, it can not be
any \emph{mixture} of these colours.
  The answer is that on a \emph{single copy} the outcome
for both the new states will be randomly red or yellow. Further, if the outcome is yellow, the state after measurement is $|yellow\rangle$
etc.
 That on an \emph{ensemble}, the \emph{probability} of each outcome is 50\%.
 This is a generalization of the famous \emph{Born probability rule} given originally in the context of \emph{collisions} \cite{bornprob}, an essential pillar of quantum theory.
 It should now be clear why no other superposition principle has the same depth as the principle of superposition of \emph{states}. 
  \subsection{Superposition principle and Heisenberg uncertainty}
 It is easy to invert the previous combinations, to get
\be
|red\rangle = \frac{1}{\sqrt{2}}\left\{|good\rangle+|bad\rangle\right\}
\ee
and
\be
 |yellow\rangle = \frac{1}{\sqrt{2}}\left\{|good\rangle-|bad\rangle\right\}
\ee
 It turns out that the states $|good\rangle\,,|bad\rangle$ have \emph{definite values} of some other
\emph{attribute} which we could call \emph{smell}!
 Suppose we start with $|good\rangle$ and make a \emph{colour} measurement. The outcome will be red or yellow with equal
probability. If it is red, the state after the measurement is $|red\rangle$, and likewise for the outcome yellow.
 Let us say that the outcome is \emph{red}.
 Now imagine a \emph{smell} measurement on the system.
 Because the state after the last measurement i.e $|red\rangle$ is an \emph{equal} superposition of the good and bad
smell states, the outcome will be one of these randomly and with equal probability.
 Therefore, even though we started with a state whose \emph{smell} was \emph{certain} i.e good, an intervening
colour measurement has completely destroyed this certainty!
 Instead, the smell information has become totally \emph{unpredictable}!
 This is the inherent \emph{indeterminacy} of quantum theory.
 This is also a demonstration that the pair of observables \emph{colour, smell} are mutually \emph{incompatible}.
 Existence of incompatible obsrvables is the essential content of the \emph{Heisenberg Uncertainty Relations}.
  \subsection{Superposition principle and complimentarity}
 The superposition principle incorporates \emph{complimentarity} in a natural way.
 To see this, note that a given state $|\psi\rangle$ can be equivalently expressed, in the context of the so called \emph{wave}
and \emph{particle} aspects as
\be
|\psi\rangle = \sum_i\,\alpha_i\,|W_i\rangle = \sum_i\,\beta_i\,|P_i\rangle
\ee
where $P_i,W_i$ are just a symbolic shorthand for the particle and wave aspects.
 Bohr's great vision lay in the recognition of the role of \emph{measurements} in quantum theory. We immediately see
the superposition principle as an embodyment of this i.e the same $|\psi\rangle$ above reveals either the $P_i$ or the $W_i$
depending on the measurement scheme whose outcomes are one or the other.
 Bohr, in his Como lecture \cite{bohrcomp}, recognized another face of complimentarity whereby states are superpositions of what may
be called \emph{classically incompatible} states. For example, in the rose atom case, the two colors are classically
distinct and mutually exclusive. The states $|good\rangle, |bad\rangle$ are indeed such superpositions of classically incompatible
states.

 In summary, we see the superposition principle as the glue, in a precisely stated manner, of the three milestones
that underlie the \emph{physical interpretation} of quantum theory, namely, the \emph{uncertainty principle,
complimentarity principle} and the \emph{probability interpretation}.
 It should be emphasized that this is so only in conjunction with suitable rules for quantum measurements. 
 Such rules are anyway essential for a consistent interpretation of the superposition principle itself.
\section{Why the \emph{rock}?}
 In addition to its central role elaborated above, it turns out that the principle is rather \emph{rigid}.
 All efforts to modify it have resulted both in serious mathematical as well as conceptual difficulties.
 From the mathematical side the chief difficulty is in finding a smooth deformation of the Hilbert Space structure.
 From the conceptual side, difficulties include apparently unphysical effects like superluminal signal propagation etc.
\cite{weinberg,polchinski,mielnik}. 
 Even the probability interpretation may itself become very fragile as a result of such modifications.
 It is therefore apt to call the principle a \emph{Rock}!

It should however be noted that many of these attempts at modifying quantum mechanics still maintain some essential features
of the present theory while modifying some other features. These failures may only be an indication that if at all, the
whole structure may have to be overhauled.
Nevertheless, from the perspectives of a satisfactory \emph{scientific method} too, it is important to find such modifications.
 It is only when one has such a broader framework that tests of quantum theory become complete.
 It is also important to have the conceptual space to handle any future empirical developments.
 
  \subsection{How did the rock enter the foundation?}
 Because of the central nature of the superposition principle, as well as its extreme allergy to any change, it is
important to understand its \emph{genesis}.
 One needs to get a historical as well as a conceptual perspective on it.
 \emph{History of an idea is as important as the idea itself}.
 In particular, it is of value to know the attitude of the founding fathers towards such a deep and central principle.
 Such an exercise is not just of historic value, it may prove critical should a need arise in the future to modify this
principle.
 With that spirit, I have tried to uncover the development of this concept during various stages of the creation of quantum theory.
I have mainly used Bohr's Como lecture \cite{bohrcomp}, the first two editions of the book by Dirac \cite{diracbook1,diracbook2},
the early papers of Dirac \cite{diracpaper1,diracpaper2,diracpaper3}, the original papers on the discovery of quantum theory \cite{qorig},
including the \emph{Bohr triology} \cite{bohrtrilogy}, the two papers by Born on the probability interpretation \cite{bornprob}, Heisenberg's
uncertainty paper \cite{heisenunc}, as well a paper by Schr\"odinger \cite{schrounc} that played an important 
role in Bohr's analysis of the uncertainty relations. Thanks to Bacciagaluppi and Valentini \cite{solvay} the proceedings of the famous
Solvay meeting in 1927 was available in english. Initially I went carefully through this very valuable source but did not find much
explicit reference to the broader nuances of the superposition principle there.
  \subsection{Early Views}
 Before undertaking this exercise, it is worthwhile to look at some early attitudes towards the superposition principle.
 As articulated by Bohr on many occasions, and as explicitly elaborated in his Como lecture \cite{bohrcomp}, the superposition principle 
was viewed essentially in a \emph{wave theory} perspective.
 This was manifest when Bohr sought to correct Heisenberg's initially flawed thinking on the uncertainty principle. Bohr,
following Schr\"odinger \cite{schrounc},
had sought to view states with sharply localized position as a \emph{superposition} of plane waves, the latter corresponding
to states of definite momentum.
 But I had earlier said that the principle of superposition of states is unlike any other superposition principle, including the
one in wave mechanics.
 The subtlety here is that the superposition of waves of different wavelengths should really be understood as a superposition
of \emph{states} and the \emph{probability interpretation} invoked to understand the outcomes of measurements. This latter element is not
present in the classical wave superposition.
 Thus it is really the superposition principle for states which is at the heart of Bohr's explanation of the uncertainty
principle on the basis of complementarity.
 In the discussions at the Solvay meeting of 1927 too (following the report of Bragg on X-ray scattering) \cite{solvay}, superposition 
principle was essentially seen as the manifestation of wave behaviour.

  \subsection{States in quantum theory}
 It is clear that in understanding the genesis of this principle, it is first necessary to understand the genesis of
the notion of a \emph{quantum state}.
 It is often implicitly understood that the superposition principle follows simply on the basis that the Schr\"odinger
equation is \emph{linear}.
 This is not quite so. All that the linearity of the Schr\"odinger equation provides is a \emph{superposition principle
for wavefunctions}.
 This does not, however, immediately translate into a superposition principle for \emph{states}.
 This is because in the early stages it was not obvious that every wavefunction should be associated with a \emph{physical state}.
 Bohr, in his Como lecture, does provide an interesting example of how superposition of the wavefunctions of two \emph{stationary} 
states also represents a
bona fide physical state.
 It should also be stressed that at the time Schr\"odinger proposed his wave equation, \emph{Solitons} (originally discovered by Scott
Russell in 1834) were well known, and there was no \emph{fundamental} justification for a \emph{linear} wave equation.
 In fact we shall see that even in Heisenberg's matrix mechanics, the nature of quantum states was obscure in the beginning.
 We shall return to these points soon.
  \subsection{The Bohr Atom: the seed of the banyan tree}
 The idea of a \emph{quantum state} can rightly be considered to have been born with the Bohr model for atoms \cite{bohrtrilogy}.
 The central concept there was that of the \emph{stationary states}.
 The states of the Bohr atom were truly schizophrenic - one part, the stationary states, looking boldly into the future,
surviving all the epoch-making developments of quantum theory. The other, the orbits, looking embarassedly into a classical
past!
 Taking the legitimate point of view that the stationary states are \emph{quantum states}, a number of fundamental questions
would arise regarding the nature of the quantum states.
 The most important of these would be \emph{"Do the stationary states represent ALL the quantum states?"}
 The Bohr theory had no ready answers to this important question.

 However, even at that stage, Bohr's \emph{correspondence principle} would have something very relevant to say.
 The 'number' of \emph{classical states} of a Keplerian system are far more than the 'number' of stationary states of the
Bohr atom.
 This means there ought to be many more quantum states than the stationary states.
 The question is, \emph{What are these additional states?}
It is clear that there was nothing even remotely like the superposition principle in the Bohr model. Even the remarkable extensions by
the Bohr-sommerfeld model did not change this situation. However, an interesting clue
in that direction was revealed by the analysis of the \emph{Stark effect} within the Bohr-Sommerfeld theory \cite{epstein}.
Starting with zero electric field in certain stationary states, turning on the electric field and eventually turning off the field
would not return the atom to the starting state. The eventual resolution of this was to be found in the superposition principle
applied to the \emph{degenerate} states of the atom in zero field \cite{epstein2}.

  \subsection{States in Heisenberg Matrix Mechanics}
 The next revolutionary development was Heisenberg's \emph{Matrix Mechanics}.
 While this retained Bohr's staionary states, it completely did away with the classical trajectories.
 Instead, it associated the classical observables like position and momentum with \emph{matrices} whose rows and
columns were labelled by the stationary states.
 Thus, as far as the question of \emph{the totality} of quantum states was concerned, the Heisenberg matrix mechanics had gone no further
than Bohr's atomic model.
 As mentioned before, Bohr, in the Como lecture had discussed how superposing the wavefunctions
of two stationary states of an atom led to a description of the physical circumstance of radiating charges.
 Though a pointer in the right direction, this was \emph{insufficient} to arrive at the principle of superposition of states
in its entirety.

 Soon after Heisenberg's paper on matrix mechanics, Dirac made an \emph{astute}, and \emph{fundamental}, observation.
 He noted that labelling the matrices only by the \emph{stationary states} was unnecessarily restrictive.
 He proposed that it was sufficient to have the observables obey the same algebraic properties of matrices and that
the labelling could be arbitrary \cite{diracpaper1,diracpaper2}.
 This observation is what really opened up the space of physical states \emph{eventually}, though not \emph{immediately}.
 This was also the proposal of Born and Jordan, and Born, Heisenberg and Jordan \cite{bornjordan}.
 But it is also this proposal that put quantum theory into the \emph{linear straitjacket}!
 The arbitrariness in the labelling of the matrices meant that vectors used for labelling can be transformed among each other
(hence the name \emph{transformation theory}).
 In the original Heisenberg formulation, these vectors corresponded to the stationary states.
 Hence they represented quantum states.
 Now the question is whether every vector into which the stationary state vectors are transformed also represented
quantum states? 
 This is analogous to the question raised previously as to whether every superposition of Schr\"odinger wavefunctions
represents physical states.

 What led to a clarification of this question, and eventually to the superposition principle, was a critical observation by
Dirac \cite{diracpaper3}. His position was based on the centrality of the \emph{eigenvalues} of the Hamiltonian(energy) in matrix mechanics, namely, that
the eigenvalues of the energy matrix in the stationary states were to be interpreted as the \emph{value} of energy in the corresponding
stationary state. What gave support to this rather mathematical conjecture was the empirical success in identifying the eigenvalues
with the \emph{terms} of the spectra of atoms. Another important property of the Hamiltonian matrix, noted in Heisenberg's matrix
mechanics paper \cite{qorig}, was that it was \emph{Hermitean}. 

Dirac's crucial clarification is to be understood in two parts. In the first part, Dirac dealt with the so called \emph{constants of motion}.
In the version of matrix mechanics elaborated by Dirac, these were \emph{constant matrices}. Extrapolating the hermiticity requirement
to these constant matrices also, Dirac made the suggestion that their eigenvalues in the stationary states are to be interpreted as the
\emph{values} of these constants of integration in the corresponding stationary states. This was consistent because the matrix mechanics
required these matrices to \emph{commute} with the Hamiltonian matrix (see eqn.(11) of $\S 2$ of \cite{diracpaper3}. Unlike the empirical
support for the eigenvalue idea for the Hamiltonian, the eigenvalue idea for observables did not have such an immediate empirical
support. The second part
consists of the reasonable extension that just as stationary states were states
with definite values of \emph{energy}, there ought to be quantum states with definite values for a set of \emph{compatible}
observables, even in the general case when these are not constants of motion.
\section{Back to the superposition principle..}
 It was then natural to consider the so called \emph{eigenstates} of any observable, say, position(or momentum), as
also \emph{physical states}.
The principle of superposition of states then follows as a consequence of the transformation theory, which had only remained as a mathematical property till then.
One thus ends up with many more physical states than the staionary states.
The irony is that now one ends up with far too many states than the totality of classical physical states!
  \subsection{Dirac and the superposition principle}
Dirac, more than anyone else, dwellt at length on the meaning of the superposition principle in quantum theory, and
commented both extensively and emphatically on it.
This is evident from his masterly exposition of the principle in the first and second editions of his classic book
on quantum mechanics \cite{diracbook1,diracbook2}.
In them, he explicitly showed the deep connection between the superposition principle and the foundational aspects
of quantum theory like indeterminacy, compatibility of observables etc..
He saw in it the fountainhead for a systematic and logical exposition of quantum theory, though he admitted that
that may not be the most convenient road to take.
He even sought to determine the dynamical laws governing quantum mechanics through their consistency with the superposition
principle.
\subsection{More of the principle..}
The ramifications of this principle in later developments of quantum theory have been astounding.
With the development of areas like high energy physics, superposition principle plays a central role in systems
with spin, isospin and other exotic quantum properties.
There are no descriptions for them with anything remotely resembling waves in three-dimensional space and superposition
understood in a wave theory context are irrelevant for them.
Nevertheless, the principle of superposition of states has given completely succesful description of them.
\emph{Entanglement}, a key ingredient in the fast developing areas of \emph{Quantum Information} and \emph{Quantum Computing},
is a direct consequence of the superposition of states.
Early sources of the conceptual challenges to quantum theory like the EPR paradox, and the Schr\"odinger-cat paradox are all
based on entanglement.
\section{Surreptitious or not?}
 Despite its centrality in quantum theory, the development of the principle of superposition of states appears to
have been driven more by reactions to various local issues. Every stage of this development, like the basis independence prompted
by the transformation theory, the legitimacy of including eigenvalues and eigenfunctions of generic observables etc. though very reasonable, 
do not seem inevitable. The entire development can be said to be more of an evolution rather than a revolution!
 This is in stark contrast to other great principles like the ones in \emph{special relativity}, \emph{general relativity}
etc.
 In short, its entry into the very foundations can be considered somewhat \emph{surreptitious}.
 Should future experiments belie the expectations of this principle, one may need the great intuitive powers of Bohr, Heisenberg,
Schr\"odinger and Born, as well the meticulous systematics of Dirac and the critical creativity of Einstein to get past the crisis!

\vskip 0.5cm
\noindent {\bf{Acknowledgments:}} I thank Finn Aaserud for giving me an opportunity to present these ideas at this
historical meeting. I also thank Jan Ambjorn for an invitation to spend a month at the Niels Bohr Institute, which enabled me
to attend this meeting. I am grateful to the Chennai Mathematical Institute and the Centre for Quantum Information and Quantum
Communication(CQIQC) of the Indian Institute of Science, Bangalore for their support. I also acknowledge support from 
Department of Science and Technology to the project IR/S2/PU- 001/2008.

\end{document}